\documentclass[a4paper]{jpconf}
\usepackage{graphicx}
\usepackage{amsmath}
\usepackage{esint}
\usepackage{array}
\usepackage{verbatim}
\usepackage{color}
\usepackage{colortbl}
\usepackage[dvipsnames]{xcolor}
\usepackage{listings}
\usepackage{epstopdf}
\usepackage{subfigure}
\usepackage{rotating}
\usepackage[percent]{overpic}
\usepackage{wrapfig}
\usepackage{hyperref}
\hypersetup{pdfborder={0 0 0}, colorlinks=true, citecolor=blue, linkcolor=blue}
\usepackage{url}
\usepackage{subcaption}
\graphicspath{{./}{figures/}}

\definecolor{mpldarkred}{rgb}{0.545098039216,0.0,0.0}
\definecolor{mplpowderblue}{rgb}{0.690196078431,0.878431372549,0.901960784314}
\definecolor{mplcadetblue}{rgb}{0.372549019608,0.619607843137,0.627450980392}
\definecolor{mplred}{rgb}{1.0,0.0,0.0}
\definecolor{mplblue}{rgb}{0.0,0.0,1.0}

\begin{document}
\title{Advancing parabolic operators in thermodynamic MHD models II: Evaluating a Practical Time Step Limit for Unconditionally Stable Methods}
\author{Ronald M. Caplan\textsuperscript{1}, Craig D. Johnston\textsuperscript{2,3}, Lars K. S. Daldoff\textsuperscript{4,3}, \\ and Jon A. Linker\textsuperscript{1}}
\address{$^1$ Predictive Science Inc., 9990 Mesa Rim Road Suite 170, San Diego, CA  92121
\\
$^2$ Department of Physics and Astronomy,
George Mason University,
Fairfax, VA 22030 
\\
$^3$ Heliophysics Science Division,
NASA Goddard Space Flight Center,
Greenbelt, MD 20771
\\
$^4$ Catholic University of America, 620 Michigan Ave., N.E. Washington, DC 20064
}
\ead{caplanr@predsci.com, craig.d.johnston@nasa.gov, lars.daldorff@nasa.gov, linkerj@predsci.com}

\begin{abstract}
Unconditionally stable time stepping schemes are useful and often practically necessary for advancing parabolic operators in multi-scale systems.  However, serious accuracy problems may emerge when taking time steps that far exceed the explicit stability limits.  In our previous work, we compared the accuracy and performance of advancing parabolic operators in a thermodynamic MHD model using an implicit method and an explicit super time-stepping (STS) method.  We found that while the STS method outperformed the implicit one with overall good results, it was not able to damp oscillatory behavior in the solution efficiently, hindering its practical use.  
In this follow-up work, we evaluate an easy-to-implement method for selecting a practical time step limit (PTL) for unconditionally stable schemes.  This time step is used to `cycle' the operator-split thermal conduction and viscosity parabolic operators.  We test the new time step with both an implicit and STS scheme for accuracy, performance, and scaling.  We find that, for our test cases here, the PTL dramatically improves the STS solution, matching or improving the solution of the original implicit scheme, while retaining most of its performance and scaling advantages.  The PTL shows promise to allow more accurate use of unconditionally stable schemes for parabolic operators and reliable use of STS methods.
\end{abstract}

\section{Introduction}
\label{sec:intro}

Thermodynamic magnetohydrodynamic (TMHD) models, like many other models of complex physical systems,  contain processes spanning widely varying time scales.  This makes numerical integration difficult, as the time steps required to describe the fastest processes can be quite small.   When using explicit time-stepping methods, the stability requirements for these fast processes can further restrict the time steps (often a great deal), making the simulations computationally intractable.  To combat this problem, unconditionally stable time-stepping algorithms can be used, effectively eliminating the numerical time step restrictions.  However, such methods are prone to miss-use, as integrating them with very large time steps can cause fast processes of the system to not be captured, and/or introduce large errors to the system. 

In our previous work (\cite{Caplan_2017}, referred henceforth as `Paper I'), we compared two popular unconditionally stable time-stepping algorithms applied to the parabolic operators within the Magnetospheric Algorithm outside a Sphere (MAS) TMHD model, which is used to study the solar corona and heliosphere (see references therein). These were the L-stable \cite{ALstable} implicit Backward Euler (BE) method solved with a preconditioned conjugate gradient (PCG) solver, and an explicit super time stepping (STS) scheme, specifically, the A-stable \cite{ALstable} second-order Legendre Extended Stability Runge-Kutta method (RKL2) \cite{RKL2_2014}.  It was found that the RKL2 method computationally outperformed BE+PCG (especially in scalability across many CPUs), but suffered from solution artifacts in some isolated regions of the model domain.  These artifacts appeared to be the result of using too large a time step combined with the poor high-mode damping of the A-stable RKL2 method when applied to the artificial kinematic viscosity term used to damp small spurious oscillations arising from other parts of the model.

This issue raised concerns about the robust use of STS methods in general, and in MAS, led to limiting its production use to a small selection of run types. While the BE+PCG's L-stable properties made it more robust (avoiding the solution artifacts), we have observed cases where it too can suffer from weak damping of oscillations when given too large of a time-step.  One way to mitigate this issue is to `cycle' the parabolic operator, effectively reducing the time-step, and allowing the high modes to damp through successive application of the operator.  However, it is difficult to know how many cycles are required a-priory \cite{DAWES2021}, and overestimation can significantly reduce the computational performance.

In this work, we follow-up Paper I by evaluating a novel, easy-to-implement method for dynamically calculating a practical time step limit (PTL) for unconditionally stable time stepping schemes applied to parabolic PDEs.  The PTL is implemented in MAS, where we use it to automatically set the minimum number of cycles needed for the operator-split parabolic operators over one full MAS time step (set by the flow CFL condition).  To improve performance, the PTL is re-evaluated dynamically each cycle, reducing the total number of cycles needed over the full time step.  As in Paper I, we test the PTL using an implicit scheme and an STS scheme.  We use two real-world cases to compare solution quality, performance, and scaling.

The paper is organized as follows. In Sec.~\ref{sec:ptl}, we describe the parabolic operators in the MAS model and present a brief derivation of the PTL.  Then, in Sec.\ref{sec:setup}, we describe the test setup, including the unconditionally stable methods, the test cases, and the computational environment.  Solutions computed with and without using the PTL are evaluated in Sec.~\ref{sec:results}, including performance and scaling results.  We conclude with a discussion in Sec.\ref{sec:disc}.

\section{A practical time step limit for integrating parabolic operators}
\label{sec:ptl}
Often, parts of a physical model that have smaller time scales than others are operator-split, such that each operator is integrated over the overall time step using the partial solution of the operator before it.  The order and time steps of the splitting can be done in several ways, each with a cost of adding a splitting error to the system \cite{OPSPLIT}.  Each split operator is advanced over the time step independently by integrating
\begin{equation}
\label{eq:par}
\frac{\partial\,{\bf u}}{\partial t} = {\bf F}({\bf u}),
\end{equation}
where ${\bf F}({\bf u})\equiv {\bf F}({\bf x},{\bf u},...)$ is the currently split operator.

The MAS model\footnote{See \url{predsci.com/mas} for additional details and references} described in Paper I (see references within) uses an overall time step set by a flow CFL condition, and operator splits several terms that would require very small time steps.  These include the parabolic operators of artificial kinematic viscosity and Spitzer thermal conduction \cite{2014_Priest_BOOK} given by
\begin{equation}
\label{eq:visc}
{\bf F}_{\mbox{\scriptsize visc}}({\bf v}) =\frac{1}{\rho}\,\nabla\cdot\left(\nu({\bf x})\,\rho\,\nabla {\bf v}\right), \; \mbox{and} \; {\bf F}_{\mbox{\scriptsize tc}}(T) =\frac{(\gamma-1)\,m_p}{2\,k\,\rho}\,\nabla\cdot\left(f_{\mbox{c}}(r)\,f_{m}(T_0)\,\kappa_0\,T_0^{5/2}\,
{\bf \hat b}{\bf \hat b}\cdot \nabla T\right),
\end{equation}
where ${\bf v}$, $T$, and $\rho$ are the plasma flow velocity, temperature, and density respectively, $\nu({\bf x})$ is the coefficient of viscosity, ${\bf \hat b}=|{\bf B}|/{\bf B}$ is the normalized direction of the magnetic field, $\gamma=5/3$ is the adiabatic index, $m_p$ is the proton mass, $k$ is Boltzman's constant, $\kappa_0$ is the Spitzer coefficient for thermal conduction, $f_{m}(T)$ is used to increase the parallel thermal conductivity at low temperatures which, in conjunction with an inverse modification to radiative cooling, broadens the transition region \cite{TCmod2001}, $f_{\mbox{c}}(r)$ is a profile that limits the radial extent that collisional thermal conduction is active, and $T_0$ is the previous step's temperature used to keep the operator linear through lagged diffusivity \cite{LAG_DIF}.

If the numerical stability time step limit of the operator-split parabolic term is smaller than the overall step, it can be cycled at its stable limit (at extreme computational cost), or an unconditionally stable scheme can be utilized.  As described in Sec.~\ref{sec:intro}, the latter can lead to problems when the overall time step is very large compared to the explicit limit.  In such cases, the operator can be cycled (with many less cycles than an explicit scheme) to reduce solution artifacts.  However, the number of cycles is typically set by experimentation (see the discussion in Ref.~\cite{DAWES2021}) which can be time consuming to determine, and is therefore not practical for large MHD models.

A novel method to compute a practical time-step limit (PTL) for a parabolic operator is introduced in Ref~\cite{PTLTHEORY}.   Given a parabolic equation of the form of Eq.~\ref{eq:par}, the goal is to maintain the sign of the difference in adjacent cells when integrating from time $t^n$ to $t^{n+1} = t^n + \Delta t$:
\begin{equation}
\label{eq:ptl}
\Delta {\bf u}^{n+1}_{\vec i}\,\Delta {\bf u}^n_{\vec i} \ge 0,
\end{equation}
where ${\vec i}$ is the grid point in question and $\Delta {\bf u}$ is the difference between the solution at ${\vec i}$ and its cell neighbor.   In general, this condition can be considered in all directions and dimensions.  For simplicity, we show one direction, where we define $\Delta {\bf u}^{n}_{i} = {\bf u}_i^{n}-{\bf u}_{i-1}^{n}$.  Using the first-order Taylor expansion of the solution ${\bf u}^{n+1}$ in $\Delta t$ and dropping higher order terms, Eq.~\ref{eq:ptl} becomes
\[
(\Delta {\bf u}^{n}_{i})^2 + \Delta t\,\Delta {\bf u}^{n}_{i}\,\Delta {\bf F}^{n}_{i} \ge 0,
\]
where $\Delta {\bf F}_{i}^n = {\bf F}_i({\bf u}^n) - {\bf F}_{i-1}({\bf u}^n)$.  The only case where this condition is not unconditionally satisfied is when $\Delta {\bf u}^{n}_{i} \ne 0$, $\Delta {\bf F}^{n}_{i} \ne 0$, and $\Delta {\bf u}^{n}_{i}\,\Delta {\bf F}^{n}_{i}<0$, in which case 
\begin{equation}
\label{eq:ptl2}
\Delta t \le -\frac{\Delta {\bf u}^{n}_{i}}{\Delta {\bf F}^{n}_{i}}.
\end{equation}
The condition can be checked in all directions surrounding grid cell $\vec i$ that meet the criteria over every grid cell, and the minimum time step calculated.  However, Ref.~\cite{PTLTHEORY} found (in their cases) that the minimum $\Delta t$ always occurred at the location of the maximum absolute value of the components of ${\bf F}$ (denoted as $F_{\mbox{\scriptsize max}}$). This allowed them to avoid needing to compute the conditions over the whole grid and sidestep numerical sensitivity issues near values very close to zero.  Here, we make the assumption that their result extends to our cases and only compute the time step conditions at the grid location of $F_{\mbox{\scriptsize max}}$, which we denote as $\vec k$. The PTL can then be defined as
\begin{equation}
\label{eq:ptl3}
\Delta t_{\mbox{\scriptsize PTL}} \le \mbox{min}\left[-\frac{u_{\vec k}^{n}-u_{\vec k+\vec i}^{n}}{F_{\vec k}(u^n) - F_{\vec k+\vec i}(u^n)}\right], 
\end{equation}
where the minimum is taken for all $\vec i$ one cell away in the 6 orthogonal grid directions.  

Once the PTL is found, the number of cycles to integrate the operator at the PTL over the larger overall time-step can be set.  However, since the operator is parabolic and smooths the solution with each application, the PTL can be dynamically re-evaluated and applied after each cycle.  This can increase the PTL size after each step, leading to much fewer overall cycles, increasing performance.  A detailed formulation and analysis of the PTL is discussed in Ref~\cite{PTLTHEORY}.  Here, we test applying the PTL to a full production code (MAS) on real-world problems.

\section{Test setup}
\label{sec:setup}

\subsection{Unconditionally stable schemes}
\label{sec:Uss}
In Paper I, we described our implementation of the first-order accurate, L-stable BE+PCG scheme, which has been successfully used in the MAS code for many years.  The implementation uses two choices of preconditioners (PC1 and PC2). PC1 (point-Jacobi) is inexpensive to formulate and apply, but limited in its effectiveness to reduce solver iterations, while PC2 (non-overlapping ILU0 factorization) is more expensive to formulate and apply, but also more effective at reducing iterations.  PC1 is easily vectorized, making it simple to implement efficiently on vector-optimized hardware (e.g. GPUs), while the standard implementation of PC2 is not vectorizable, requiring the use of alternative algorithms and libraries for use on GPUs \cite{better_than_cusparse1,better_than_cusparse2}.  The current version of MAS only supports PC1 on GPUs, while PC2 is used when running on CPUs (an important factor in comparing MAS's CPU and GPU performance).  The BE+PCG has two important drawbacks - the requirement of global communications for the dot products which hinders scaling, and the necessity of using a linear operator.

The other scheme described and tested in Paper I was the explicit second-order extended stability Runge-Kutta Legendre super time stepping (RKL2) scheme \cite{RKL2_2012,RKL2_2014}.  RKL2 is easy to implement, second-order accurate, can handle a non-linear operator, has no global synchronization points, and is vectorizable yielding efficient implementation on GPUs.  However, it is only A-stable and does not damp high modes well for large time steps.  

\begin{figure}[htbp]
\centering
\includegraphics[height=1.275in]{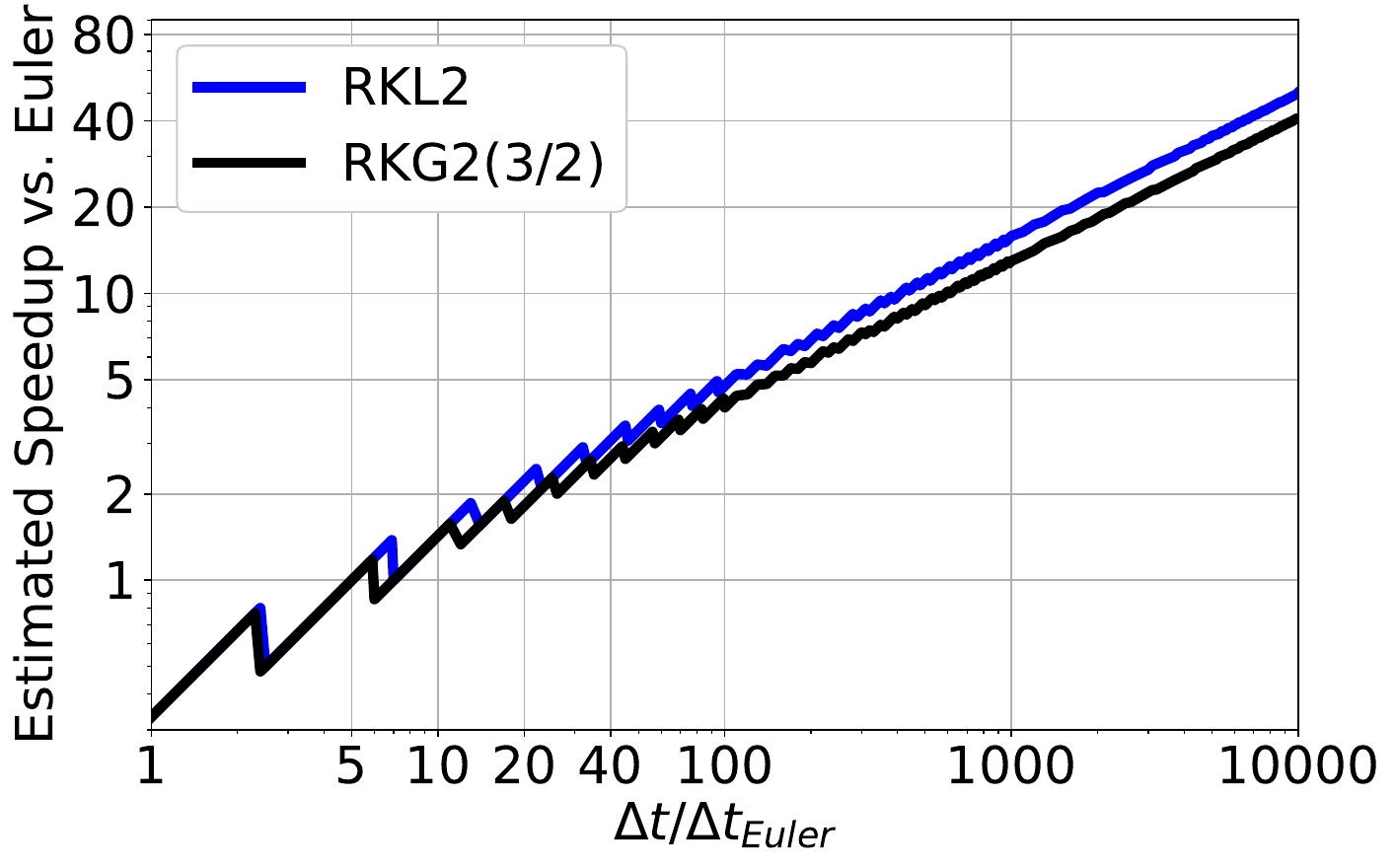}
\includegraphics[height=1.3in]{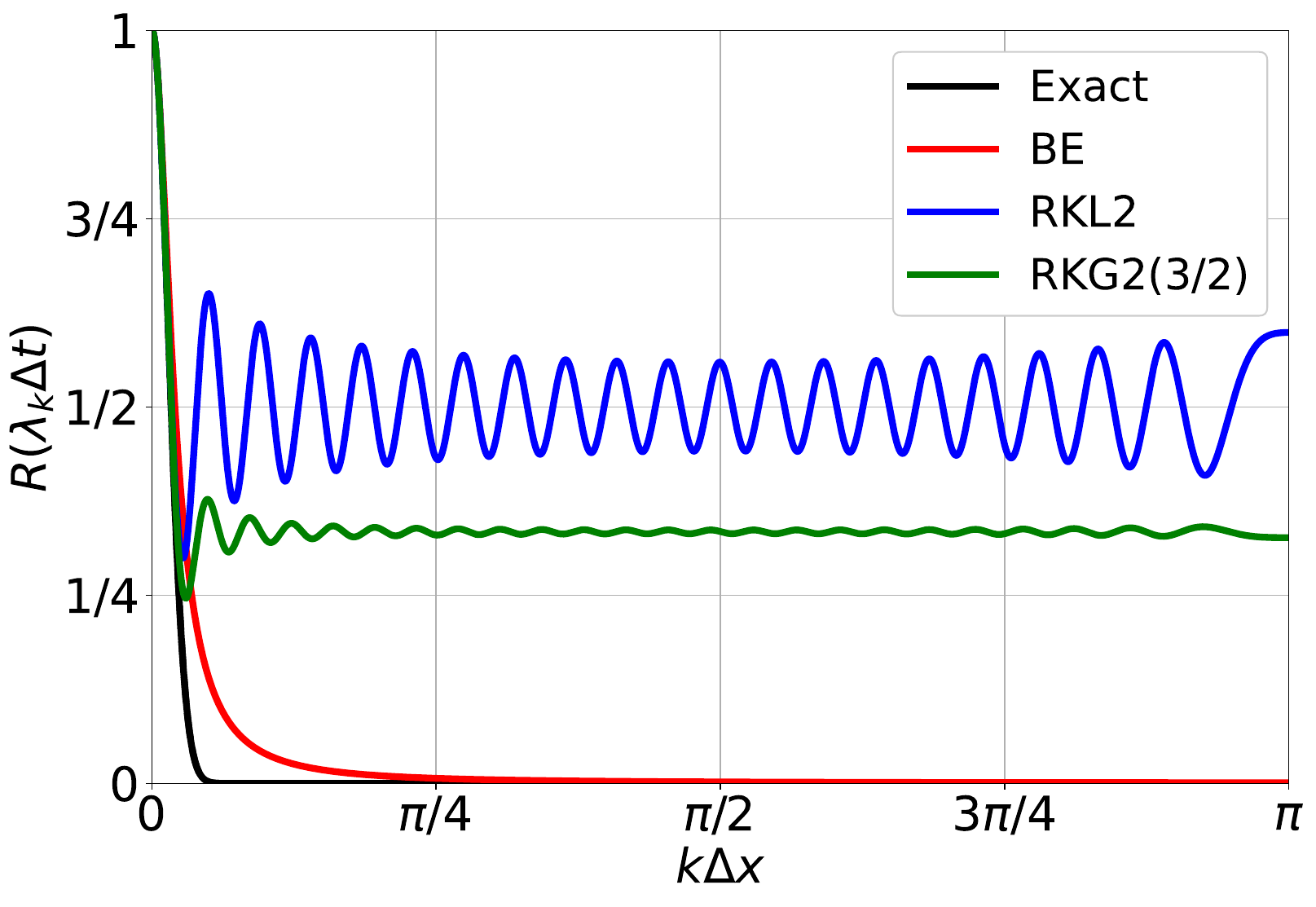} 
\includegraphics[height=1.3in]{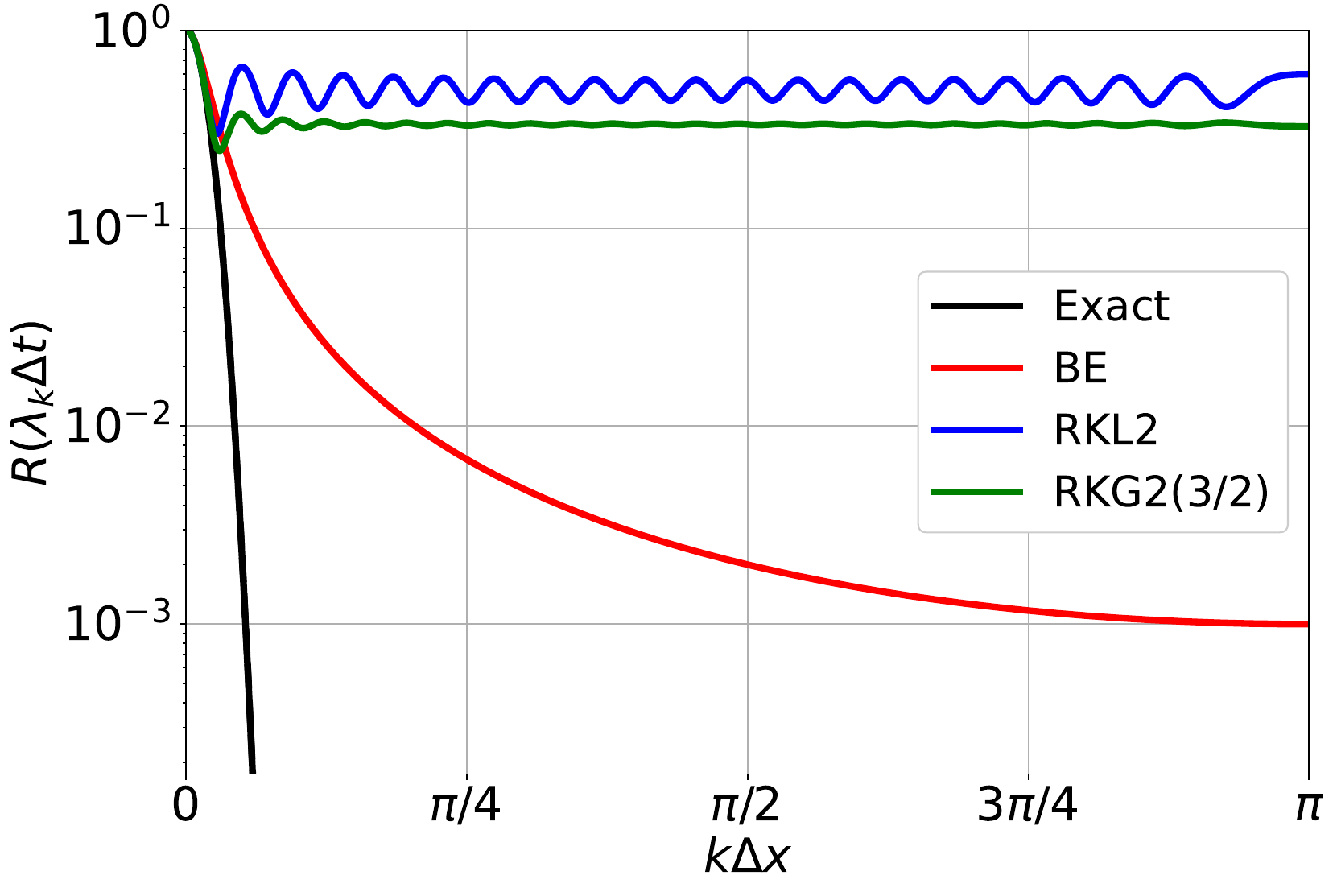} 
\caption{Left:  Estimated speedup of the RKL2 and RKG2(3/2) schemes compared to explicit Euler (approximating one Euler step to have the same computational cost as one STS iteration).  Middle: Amplification factors for the BE, RKL2, and RKG2 schemes when applied to a 1D uniform grid heat equation discretized with a second-order central finite difference at a time-step 500 times larger than the explicit Euler limit. Right: Same as middle, but on a log scale to show the difference in the BE scheme's damping rate compared to the exact solution.\label{fig:sts}}
\end{figure}

The second-order Gegenbauer method (RKG2) recently introduced in Ref.~\cite{o2019runge} has similar performance to the RKL2 scheme, but with a better damping amplification factor for high modes.  In Fig.~\ref{fig:sts}, we show the estimated speedups of RKL2 and RKG2, and compare the amplification factors of the BE, RKL2, and RKG2 with $\alpha=3/2$ (see Ref.~\cite{skaras2021super}) schemes to the exact solution of a simple 1D heat equation problem.
We see that the RKG2 scheme's amplification factor is qualitatively similar to the RKL2, but saturates to a lower value and is less variable across modes.  The RKG2 is also shown to be similar in performance to RKL2. We note that even the BE scheme's damping of modes diverges from the exact solution greatly, but overall it is much better than that of the STS schemes.  

Since the inability to efficiently damp high modes is a key factor in the solution artifacts we observe in Paper I, we choose to use RKG2 over RKL2 in this work.  We also note that it is critically important to ensure that the number of STS iterations of the RKG2 method be odd, otherwise the amplification factor approaches one at the highest modes.  Details of the RKG2 scheme implementation are shown in \ref{app:rkg2}.

\subsection{Test cases}
\label{sec:tests} 
To test the effects and performance of the PTL, we utilize two production MAS simulation runs.  The first test (Test 1) is a lower resolution version of a thermodynamic MHD relaxation based on simulations used in Ref.~\cite{reeves2019exploring}.  This test has 3.4 million highly non-uniform grid points and uses a constructed lower radial magnetic field boundary combining a global dipole with a localized bi-pole representing an active region.   The second test (Test 2) is a slightly modified (small changes in the viscosity values) version of the thermodynamic MHD relaxation used in Paper I with a size of 22.8 million grid points.  This test uses real-world observed surface radial magnetic fields for the lower boundary.  

For comparing solutions with and without using the PTL, we run each test's relaxation for a total of 8 hours of simulated physical time.  For performance and scaling results, we only use Test 2, where we start the simulation with the final output of an 8-hour relaxation, and continue the run for another 6 simulated physical minutes.  Since these performance runs are much shorter simulations than those used in  production, we subtract off the loading time of the relaxation solution initial condition from the wall clock time.

\subsection{Computational environment}
\label{sec:testcompspecs}
Since Paper I, the MAS code has been updated to run on NVIDIA GPUs through the use of OpenACC \cite{MASGPUACC} and Fortran's `do concurrent' standard parallelism constructs \cite{MASGPUDC}.  We therefore test the methods on both CPU and GPU systems.  For the solution comparison run of Test 1, we utilize an in-house workstation with an NVIDIA RTX 3090Ti GPU, while for Test 2 we use 32 nodes of the Expanse system at the San Diego Supercomputer Center (SDSC). For testing the parallel scaling performance on CPUs, we use between one and 32 nodes of Expanse, while on GPUs, we use between one and eight GPUs on a single node at the Delta system at the National Center for Supercomputing Applications (NCSA).  The details of the hardware and software configurations used are shown in \ref{app:comp}.

We note that the MAS code is highly memory-bandwidth bound, making the maximum memory bandwidth the best indication of expected relative performance between systems, given the same algorithm.  However, as discussed in Sec.~\ref{sec:Uss}, MAS uses a more effective preconditioner when run on CPUs than on GPUs, making the relative performance between CPU and GPU a practical time-to-solution comparison, and not an apples to apples hardware comparison.

%%%%%%%%%%%%%%%%%%%%%%%%%%%%%%
% RESULTS
%%%%%%%%%%%%%%%%%%%%%%%%%%%%%%

\section{Results}
\label{sec:results}

The key questions we want to address are:
\begin{enumerate}
    \item Can using the PTL with the RKG2 scheme obtain a solution similar (or better) to the original BE+PCG scheme?
    \item Does using the PTL with the BE+PCG scheme improve \emph{its} solution?
    \item How does the PTL affect performance for both BE+PCG and RKG2 schemes? 
    \item Does using the PTL allow RKG2 to be competitive with the original BE+PCG scheme in both solution quality and performance?
\end{enumerate}

\subsection{Solution comparisons}
\label{sec:results_sol}
\begin{figure}[htbp]
\centering
$\begin{array}{cccc}
{\hbox{\includegraphics[width=1.435in]{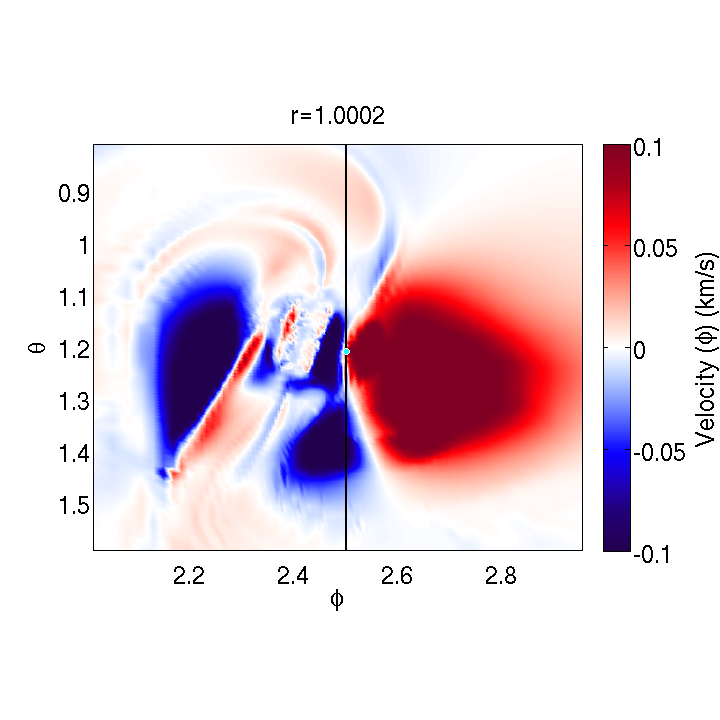}}} &
{\hbox{\includegraphics[width=1.435in]{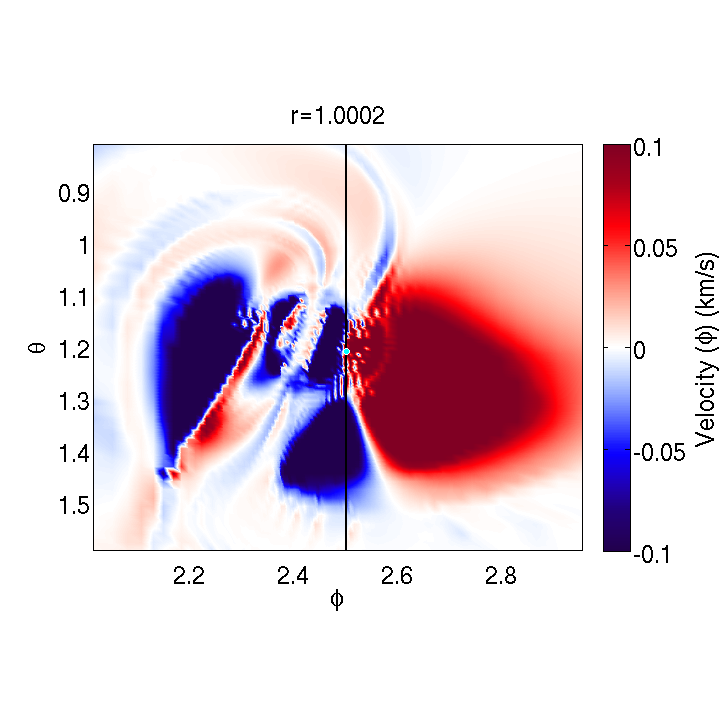}}} &
{\hbox{\includegraphics[width=1.435in]{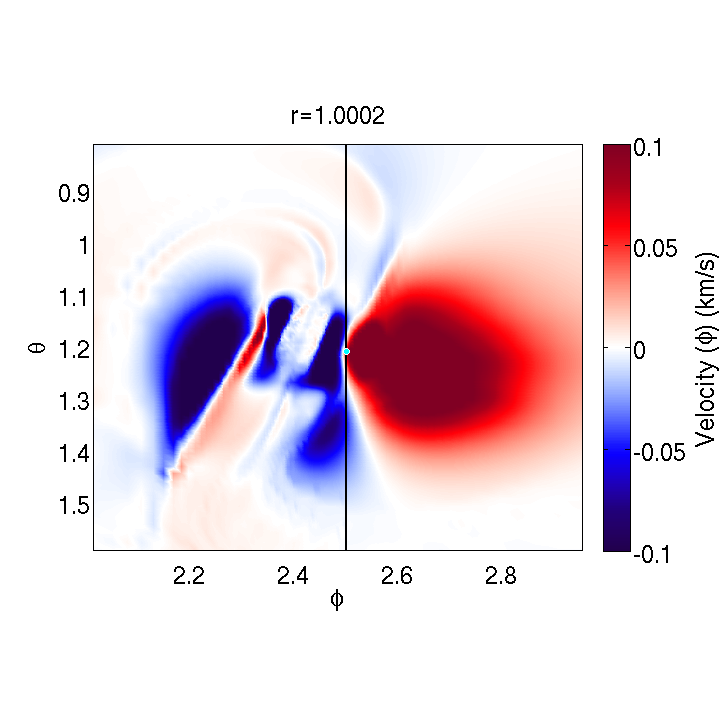}}} &
{\hbox{\includegraphics[width=1.435in]{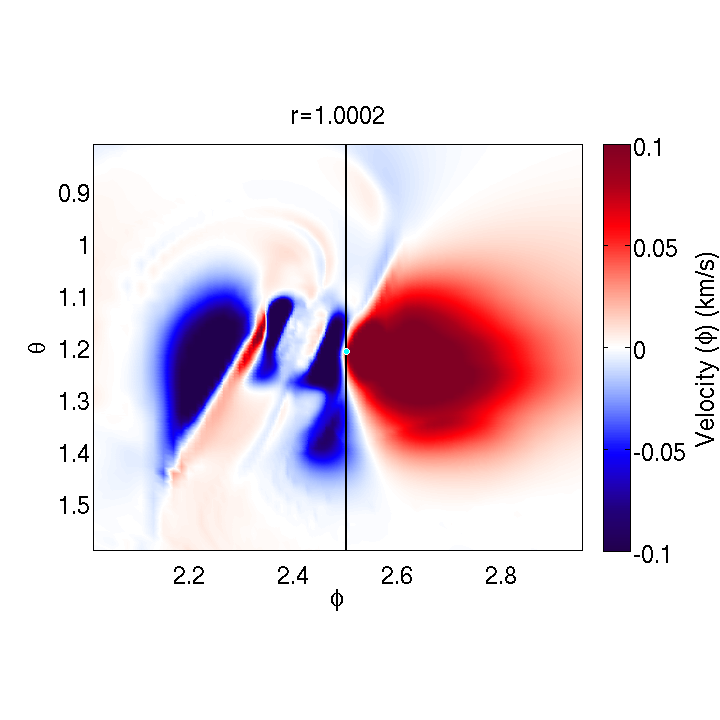}}} \\
{\hbox{\includegraphics[width=1.435in]{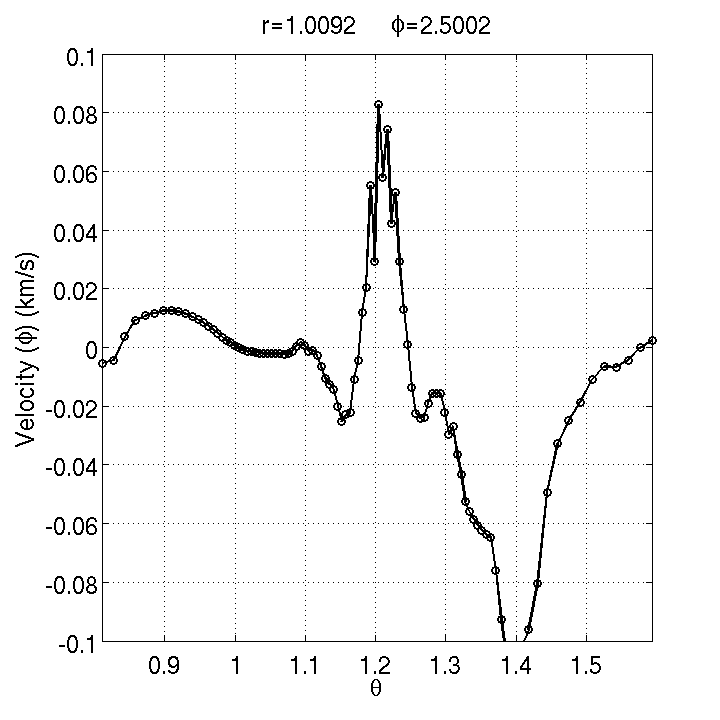}}} &
{\hbox{\includegraphics[width=1.435in]{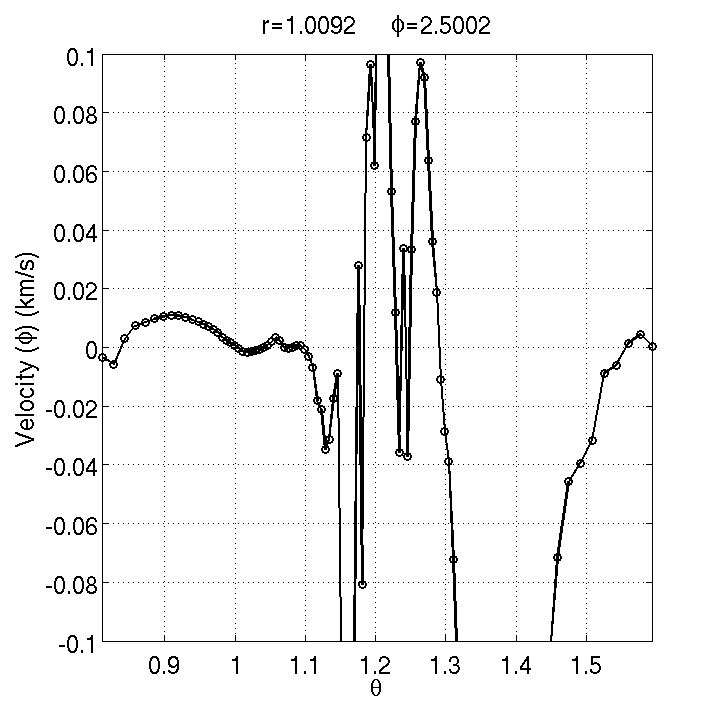}}} &
{\hbox{\includegraphics[width=1.435in]{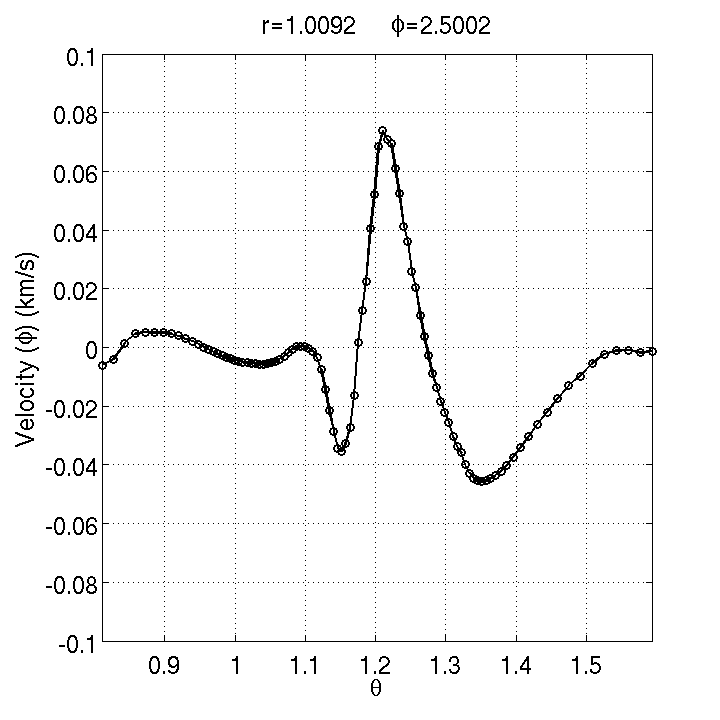}}} &
{\hbox{\includegraphics[width=1.435in]{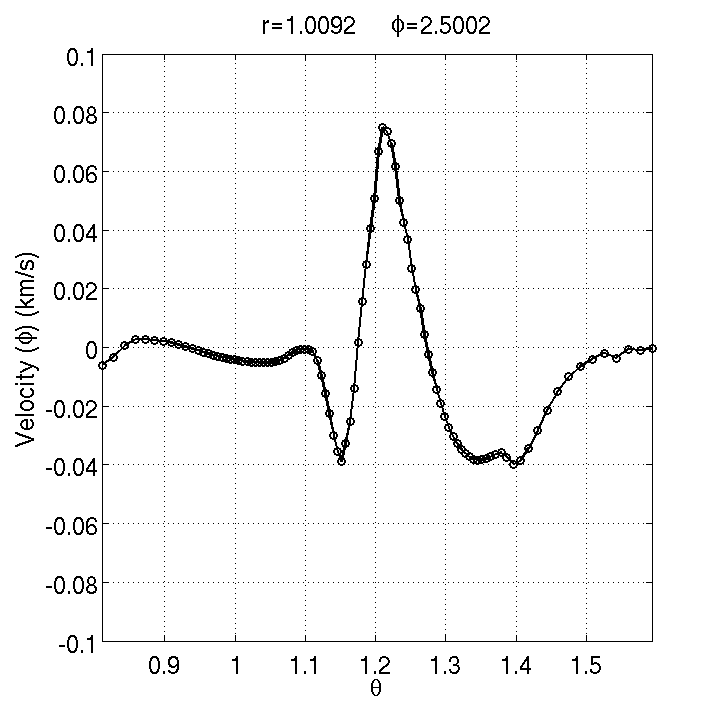}}} \\
\mbox{\color{mpldarkred} BE+PCG: Original} & \mbox{\color{mplcadetblue} RKG2: Original} & \mbox{\color{mplred} BE+PCG: PTL} & \mbox{\color{mplblue} RKG2: PTL} \\
\end{array}$
\caption{Comparison run results for Test 1.  A small section of the solution output of $v_{\phi}$ is shown in the $\theta-\phi$ plane (top) with the black line indicating the cut that is plotted in 1D (bottom).  The PTL is shown to improve the quality of the solution when using either RKG2 and BE+PCG.  The scheme used for each result is shown at the bottom of each column, where the text color corresponds to the performance result figures in the next section. 
\label{fig:test1sol}}
\end{figure}

To address the first two questions, we run the test cases with both BE+PCG and RKG2, each with and without using the PTL and compare the solutions.  Like in Paper I, the overall global solution looks qualitatively the same using all methods (not shown here).  Our focus here is on the locations and quantities within the solutions that exhibit oscillatory behavior, such as the zoomed views in Paper I's Fig.~7.  For Test 1, this issue can be seen in the $\phi$-component of the velocity field near the solar surface.  A zoomed portion of the solution, along with 1D cuts through the view are shown in Fig.~\ref{fig:test1sol}.

In this case, we notice that even the original BE+PCG scheme does not fully damp out the run's oscillations, implying that either the viscosity coefficient of the problem is not high enough, or that the time step is so large, not even the BE+PCG scheme can apply the specified viscosity correctly.  The latter seems to be the case since the PTL run with BE+PCG eliminates the oscillations, answering question (ii) in the affirmative.

For the RKG2 scheme, we see that the original implementation does a poor job applying the viscosity to damp oscillations and/or is adding its own oscillations due to the large time step (see Ref.~\cite{PTLTHEORY}).  In contrast, when using the PTL with RKG2, the oscillations have been damped, and the resulting solution is qualitatively similar to the BE+PCG solution with PTL, with only a small `notch' near $\theta=1.4$ being a noticeable difference.  The quality improvement of the solution when using the PTL is also seen clearly in the 2D cut planes, where the PTL has drastically cleaned up the original RKG2 solution.  In this case, we have shown that the answer to question (i) is `yes', as the PTL RKG2 is not only `as good' as the original BE+PCG, but, in this case, is substantially better.

\begin{figure}[htbp]
\centering
$\begin{array}{cccc}
{\hbox{\includegraphics[width=1.435in]{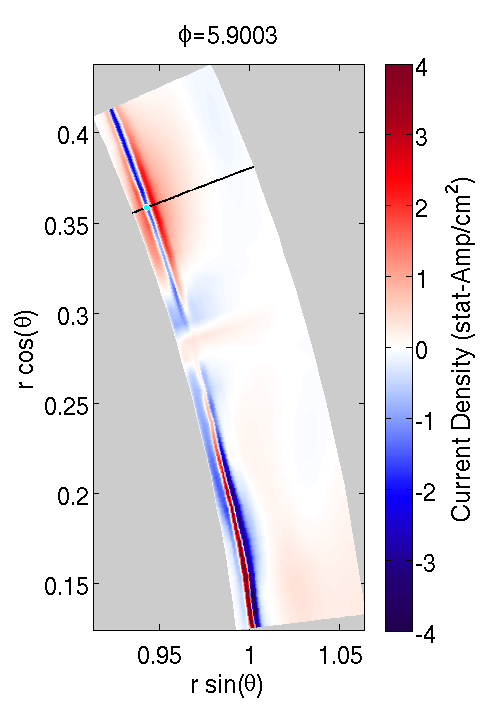}}} &
{\hbox{\includegraphics[width=1.435in]{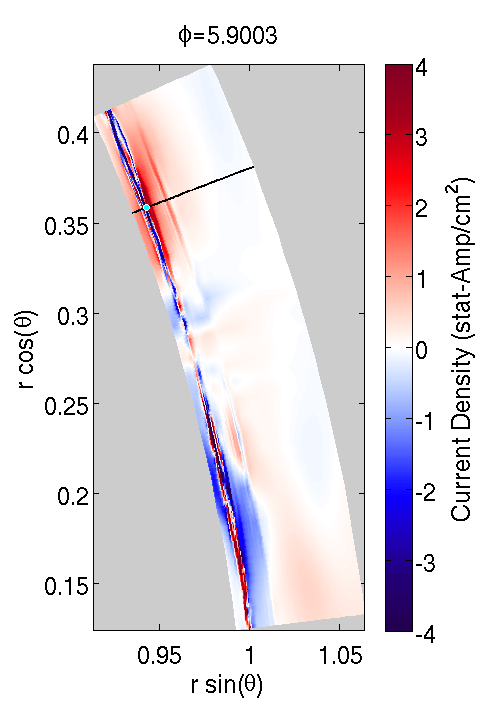}}} &
{\hbox{\includegraphics[width=1.435in]{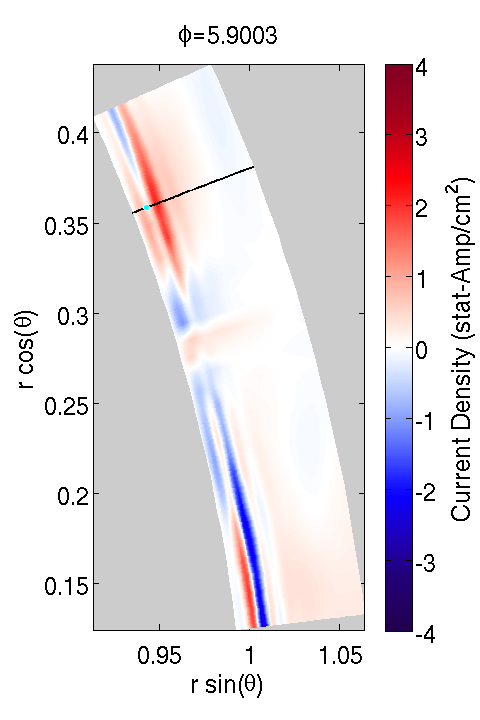}}} &
{\hbox{\includegraphics[width=1.435in]{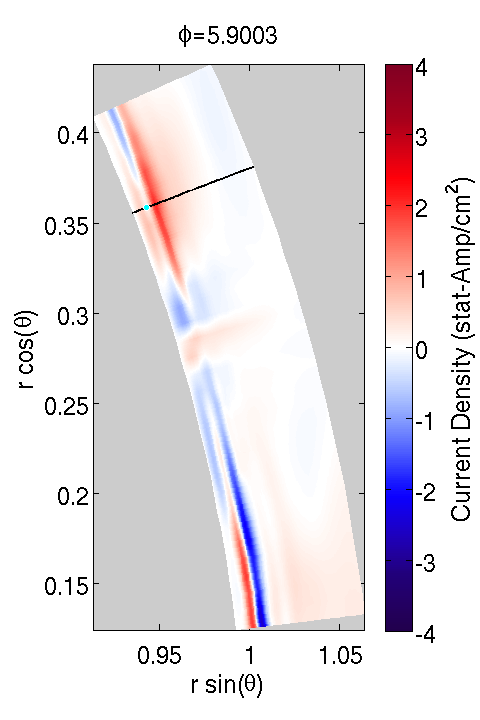}}} \\
{\hbox{\includegraphics[width=1.435in]{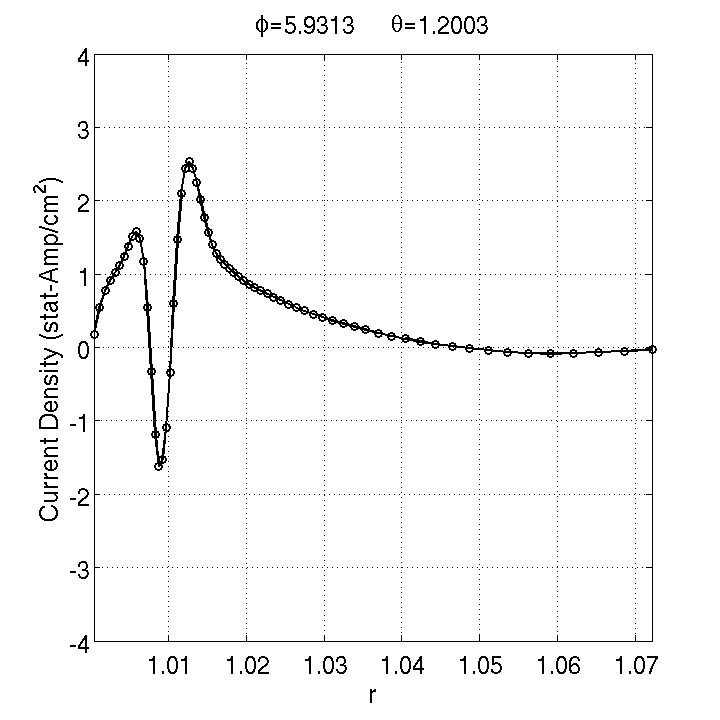}}} &
{\hbox{\includegraphics[width=1.435in]{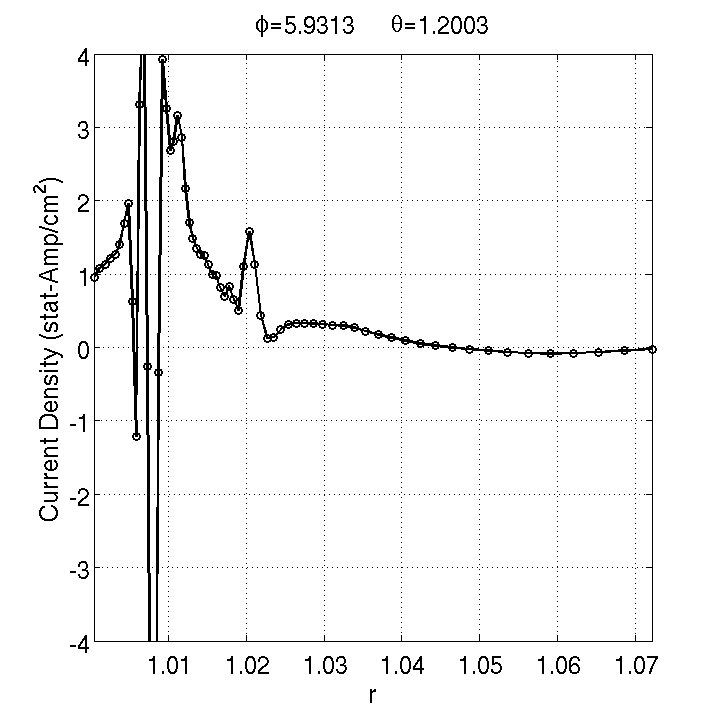}}} &
{\hbox{\includegraphics[width=1.435in]{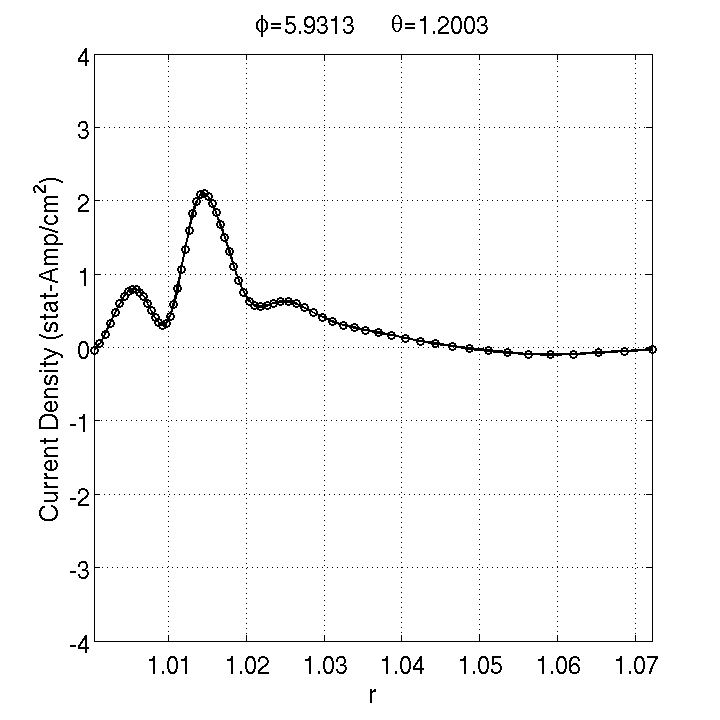}}} &
{\hbox{\includegraphics[width=1.435in]{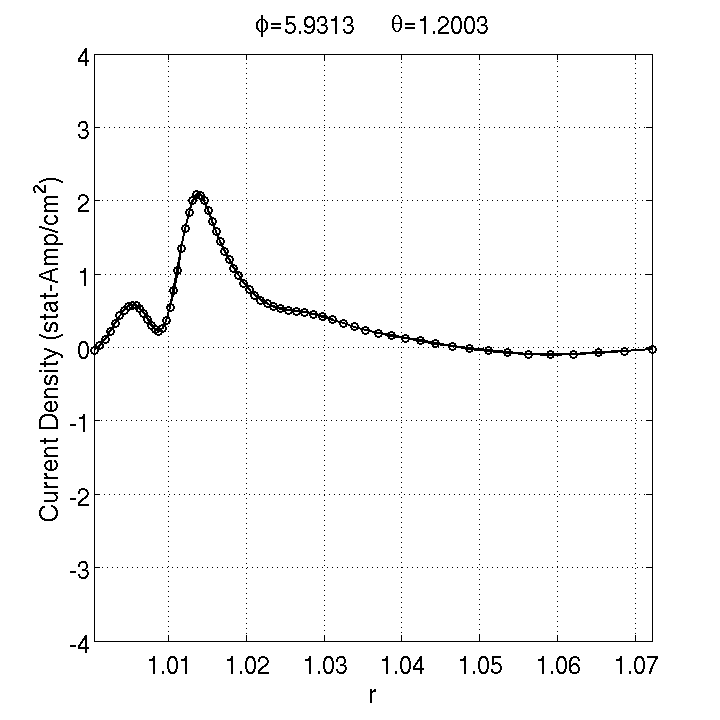}}} \\
\mbox{\color{mpldarkred} BE+PCG: Original} & \mbox{\color{mplcadetblue} RKG2: Original} & \mbox{\color{mplred} BE+PCG: PTL} & \mbox{\color{mplblue} RKG2: PTL} \\
\end{array}$
\caption{Comparison run results for Test 2.  A small section of the solution output of $j_{\phi}$ is shown in the $r-\theta$ plane (top) with the black line indicating the cut that is plotted in 1D (bottom).  The PTL is shown to improve the quality of the solution when using both RKG2 and BE+PCG.  Notation is the same as Figure~\ref{fig:test1sol}. 
\label{fig:test2sol}}
\end{figure}

Moving to Test 2, in Fig.~\ref{fig:test2sol} we show zoomed-in portions of the $\phi$-component of the current density for each solution, along with 1D cuts through the view. We see similar results as in Test 1.  The original BE+PCG scheme does not have much oscillatory behavior in this case, but when using the PTL, the solution changes and becomes much smoother overall.  The original RKG2 scheme once again suffers from oscillatory behavior, but is cleaned up substantially when applying the PTL.  In this case, the PTL RKG2 solution is even closer qualitatively to the PTL BE+PCG than in Test 1.

We can therefore conclude (at least for these cases) that the answer to both questions (i) and (ii) is `yes', as the PTL improves the solution of both schemes, and allows the RKG2 scheme to yield a very close solution to BE+PCG, with the PTL RKG2 solution being a significant improvement over the original BR+PCG scheme.  However, to know whether it is practical to use the PTL for either scheme, these solution results need to be viewed in conjunction with each scheme's computational performance.

\subsection{Performance}
\label{sec:results_perf}
\begin{figure}[htbp]
\centering
\includegraphics[width=0.495\textwidth]{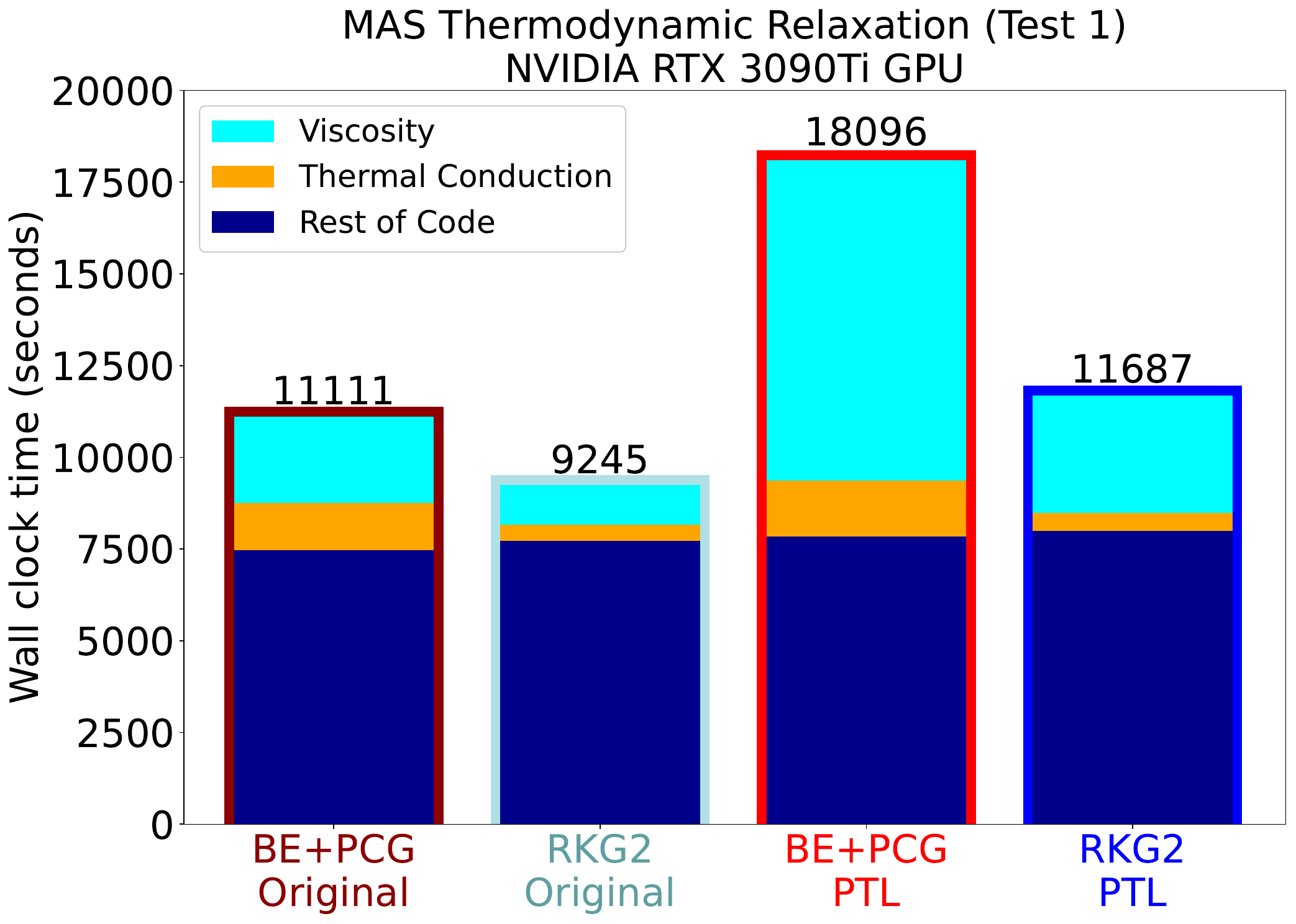}
\includegraphics[width=0.495\textwidth]{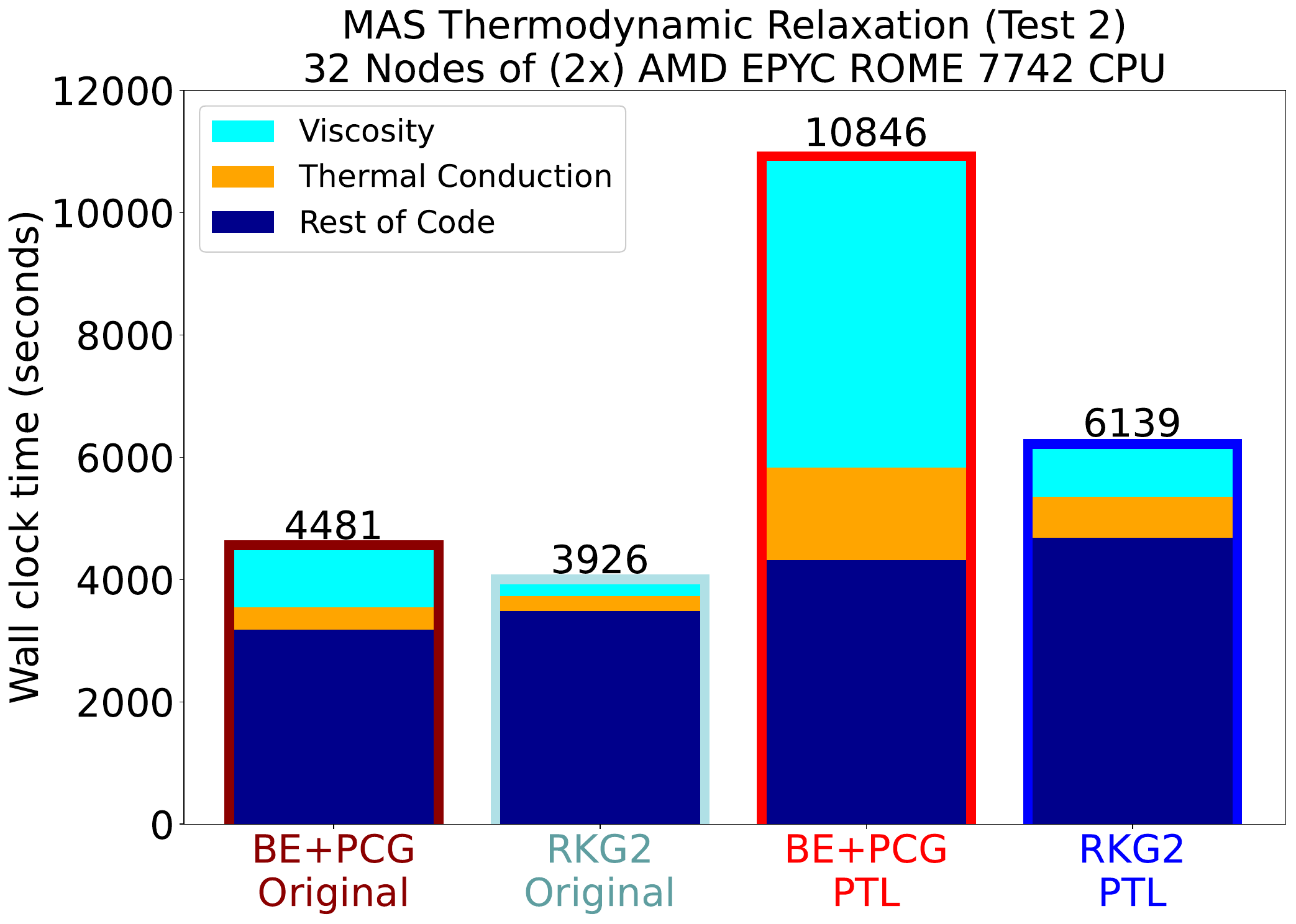}
\caption{Timing results for the solution comparison runs in Sec.~\ref{sec:results_sol} for Test 1 (left) and Test 2 (right).\label{fig:perf1}}
\end{figure}

We start by looking at the performance results for the solution comparison runs shown in the previous section, which we show in Fig.~\ref{fig:perf1}.
In both tests, we see that the original RKG2 scheme is faster than the original BE+PCG scheme as expected.  Using the PTL on BE+PCG makes the code slower by up to a factor of two, with the slowdown for viscosity being a factor of around four.  While this method did improve the solution over the original scheme, the improvement may not be worth the performance hit.  Looking at the PTL RKG2 runs, we see that, while they are slower than the original RKG2 runs, they are comparable to the run time of the original BE+PCG scheme.  Since the solution for the PTL RKG2 scheme is an improvement to the original BE+PCG solution, and comparable to the PTL BE+PCG scheme's solution, the small increase in run time is acceptable, and makes the RKG2 competitive with the original BE+PCG scheme, answering question (iv) in the affirmative for this case.

\begin{figure}[htbp]
\centering
\includegraphics[height=0.445\textwidth]{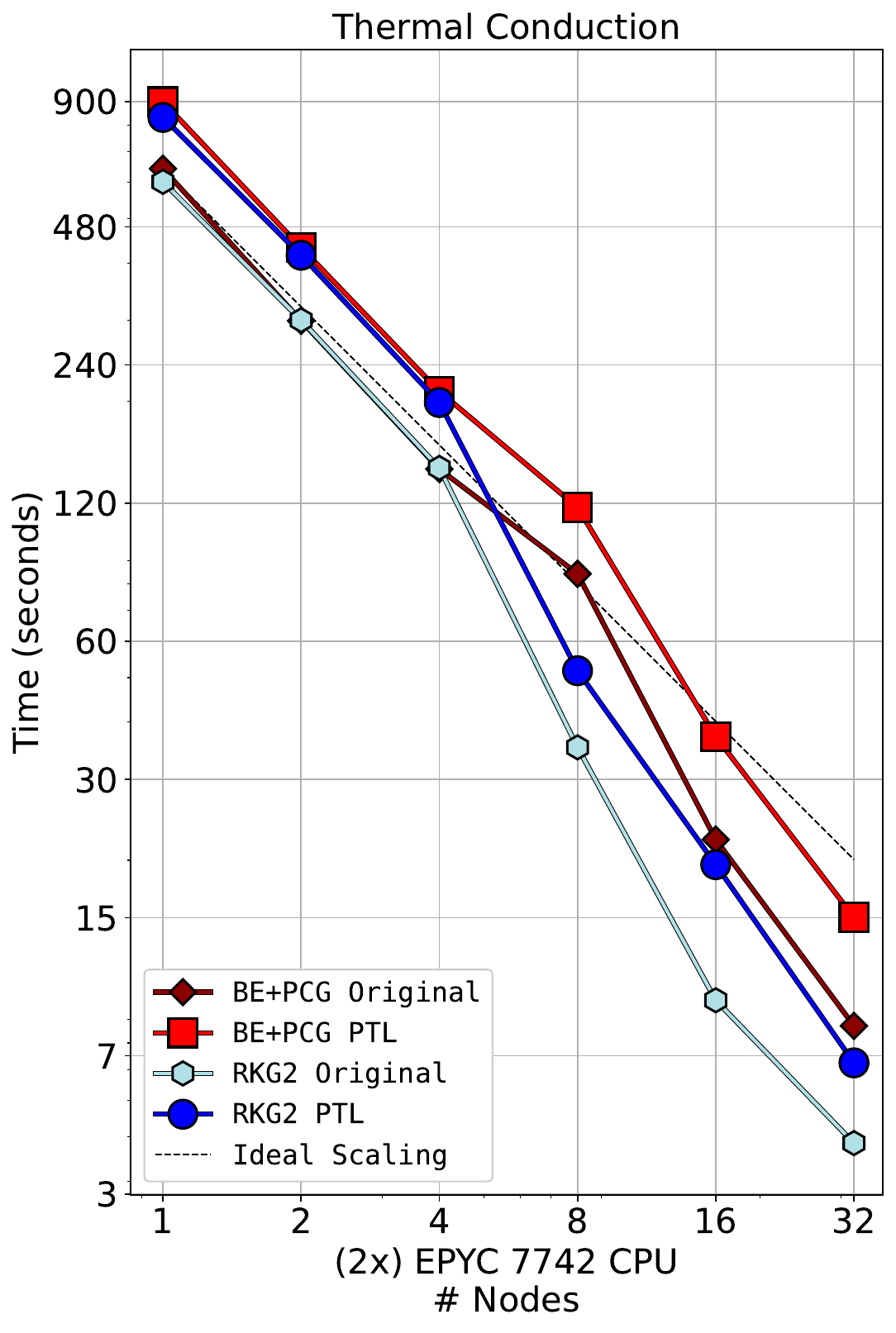}
\includegraphics[height=0.445\textwidth]{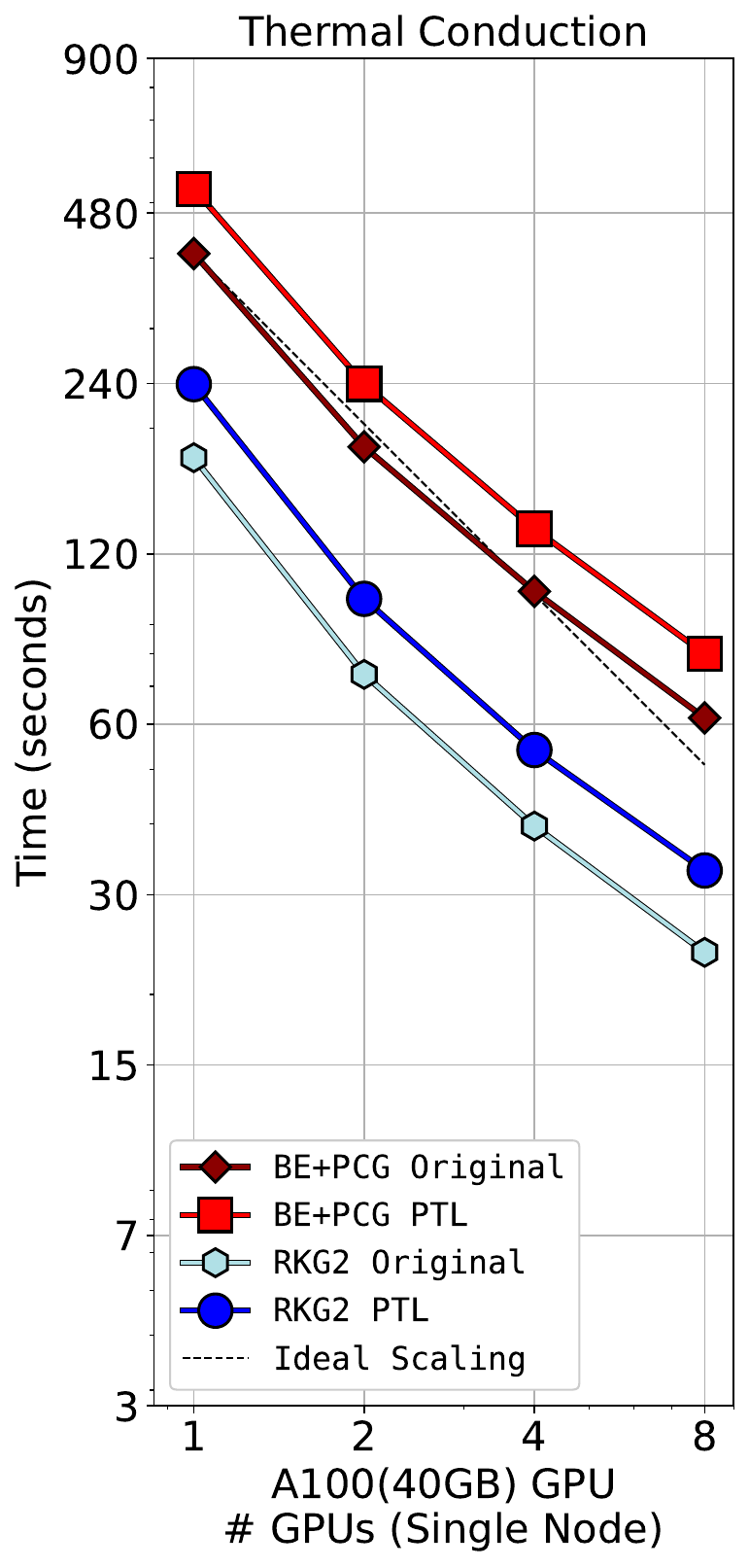}
\includegraphics[height=0.445\textwidth]{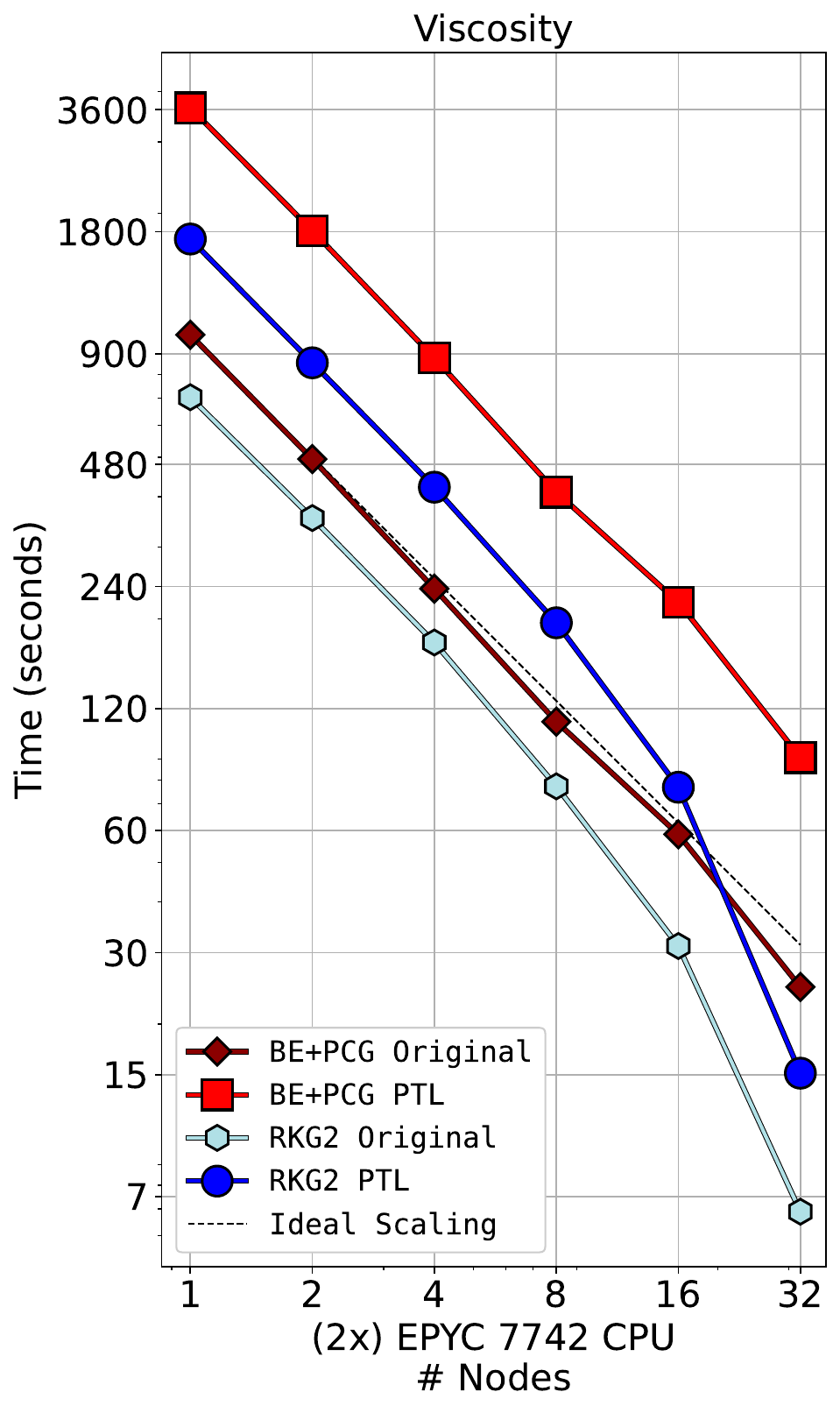}
\includegraphics[height=0.445\textwidth]{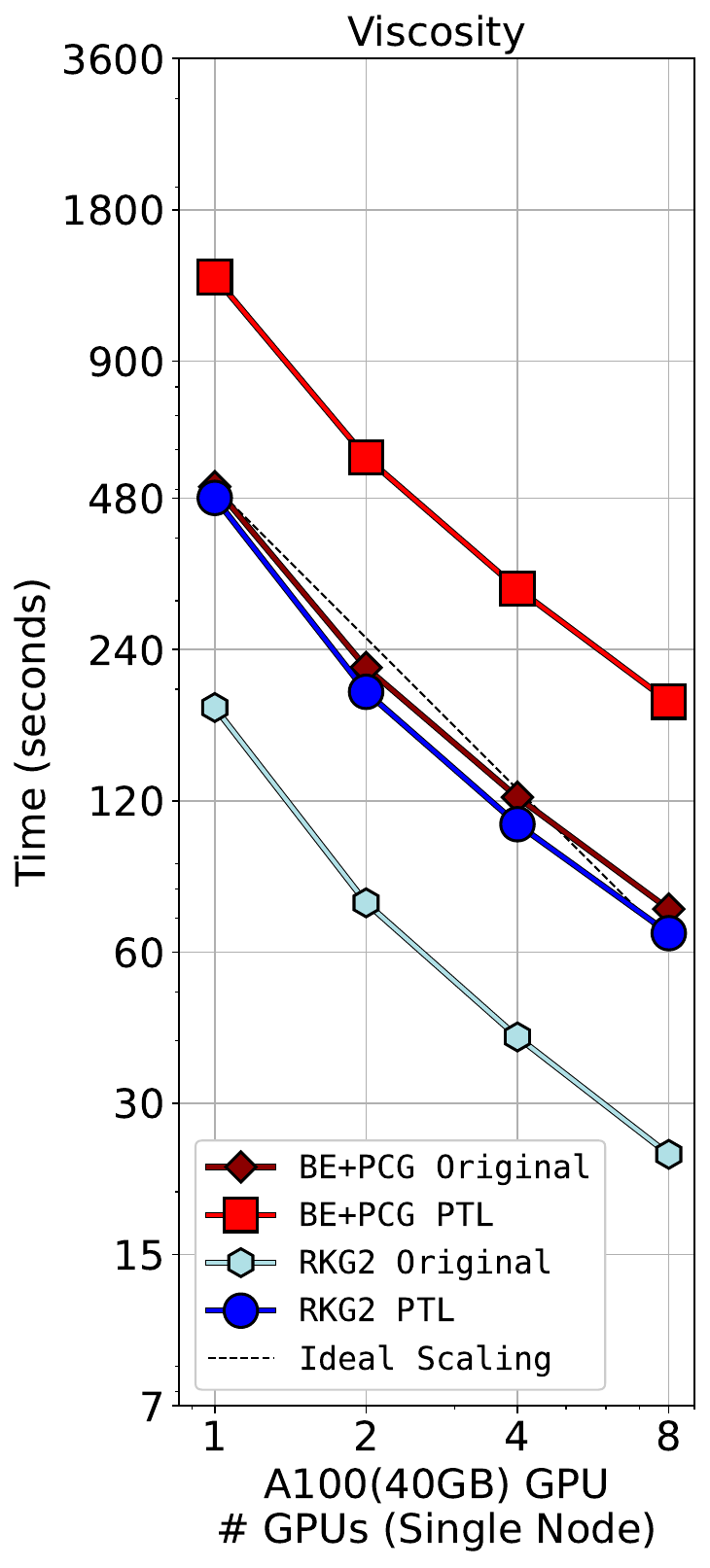}
\caption{Test 2 scaling results for the thermal conduction (left two plots) and viscosity (right two plots) operators on CPUs (left) and GPUs (right).\label{fig:perf2}}
\end{figure}

In Paper I, the performance advantages of the STS scheme over BE+PCG were even more pronounced when looking at parallel scaling across multiple CPUs.  Here, we test the scaling with and without the PTL to see if the analysis above remains valid, and if the RKG2's advantage increases.  The timing results for the two parabolic operators (thermal conduction and viscosity) portion of the runs are shown in Fig.~\ref{fig:perf2}.
On the CPUs, we see that the BE+PCG exhibits expected scaling, and for the thermal conduction operator on large numbers of nodes, exhibits `super-scaling'.  This super scaling is likely due to the very large size and speed of the EPYC CPU's cache, such that the local portion of the grid (which decreases with increasing number of CPU cores) is fitting into the fast cache. The RKG2 exhibits stronger super scaling for both the thermal conduction and viscosity operators, and is overall faster than the BR+PCG.  When using the PTL, each method runs slower than their original implementations as expected, up to a factor of four.  However, while the RKG2 with PTL runs slower than the original non-PTL BE+PCG method on small numbers of CPU nodes, when run on many CPU nodes, the super scaling causes it to be faster.  Therefore, we see that the PTL RKG2 scheme can outperform the original BE+PCG scheme, while producing a better solution as shown in Sec.~\ref{sec:results_sol}.

On the GPUs, all methods do not scale perfectly, likely caused by GPU-GPU communication overheads and the problem size not being large enough to fully saturate the GPUs.  In this case, the RKG2 is always faster than the BE+PCG.  This is expected since the GPU runs only use the less effective PC1 preconditoner for BE+PCG.  For thermal conduction, the PTL RKG2 method outperforms the original BE+PCG method by a factor of 2, while for viscosity the two run times are almost the same.  Given the superior solution quality of the PTL RKG2 over the original BE+PCG method, these results strongly favor the use of PTL with RKG2 over BE+PCG.

In all the cases shown here, we see that using PTL with BE+PCG slows down the operator considerably, by up to a factor of 4. Therefore, although the PTL with BE+PCG yields a better solution than without PTL, the drop in performance may be unacceptable.   This implies that the PTL has made STS methods competitive with BE+PCG, answering question (iv) in the affirmative. 

\begin{figure}[htbp]
\centering
\includegraphics[height=0.445\textwidth]{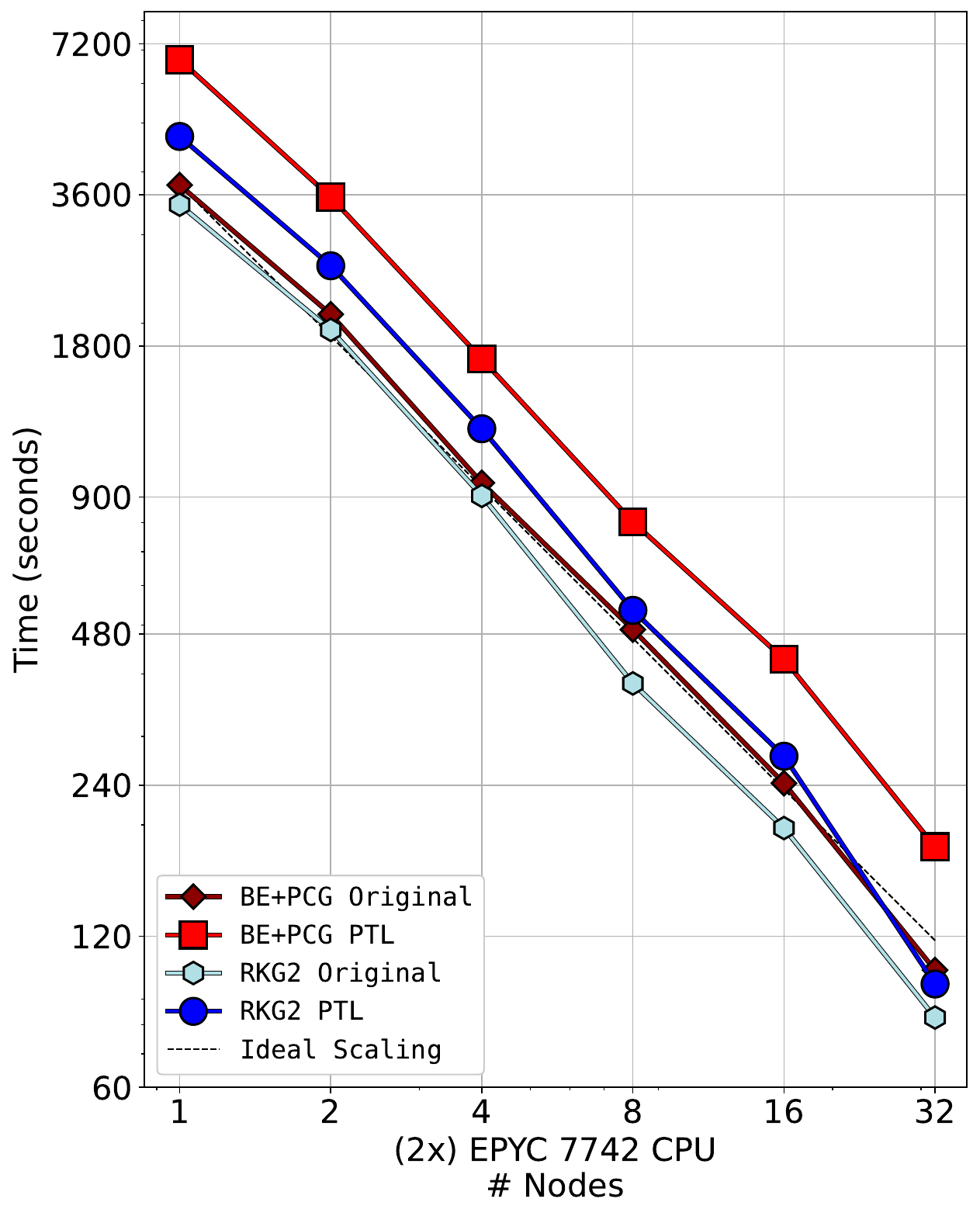}
\includegraphics[height=0.445\textwidth]{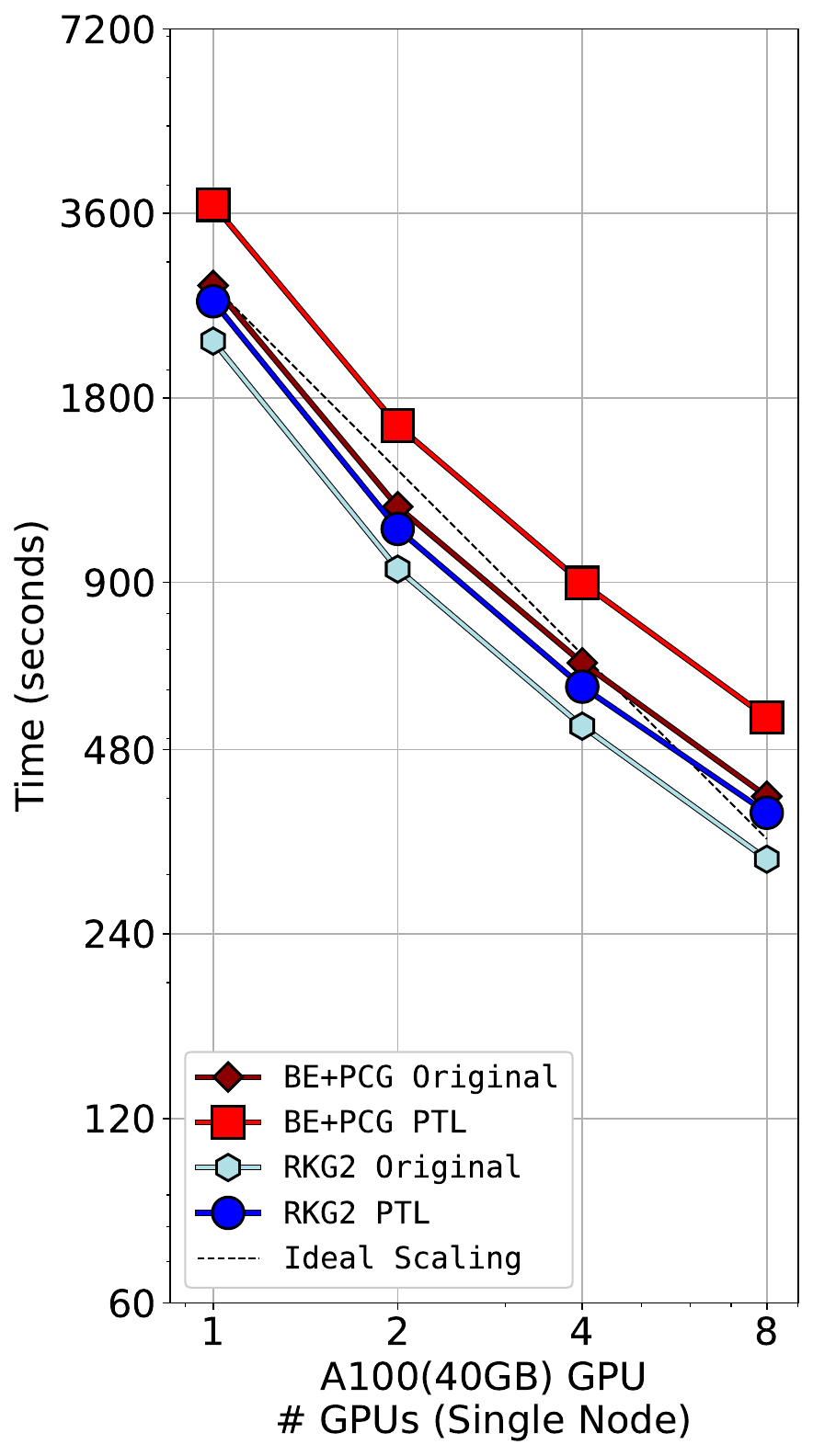}
\caption{Test 2 scaling results for total Wall clock time (with startup time removed) on CPUs (left) and GPUs (right).\label{fig:perf3}}
\end{figure}

The above analysis was for the two parabolic operators in isolation.  However, since these operators are only part of the overall run time (around $30\%$, see Paper I) of the MAS code, we also look at the effects on the total wall clock time, which we show in Fig.~\ref{fig:perf3}.
In the CPU case, the overall code scales ideally, with a small amount of super-scaling when run on 32 CPU nodes.  Given the amount of super scaling shown in Fig.~\ref{fig:perf2}, this shows that different sections of the code have varying levels of scaling efficiency.

We once again see that using the RKG2 scheme yields the best performance overall.  The RKG2 scheme with PTL is slower than the original BE+PCG scheme for small numbers of CPU nodes, but has similar performance for more than 8 CPU nodes, and surpasses its performance for 32 CPU nodes.  Using PTL with BE+PCG is once again much slower.  On GPUs, the RKG2 with PTL closely matches the performance of the original BE+PCG scheme on all numbers of GPUs.  Given the improvements of the solution using the PTL with RKG2 over the original RKG2 and BE+PCG, the new scheme is overall advantageous, where the performance loss on low numbers of CPU nodes is likely acceptable.

%%%%%%%%%%%%%%%%%%%%%%%%%%%%%%
% Discussion
%%%%%%%%%%%%%%%%%%%%%%%%%%%%%%

\section{Discussion}
\label{sec:disc}
In this paper, we have followed up our analysis in Paper I comparing unconditionally stable explicit STS schemes to implicit schemes for parabolic operators in a thermodynamic MHD model by applying an easy-to-use practical time step limit (PTL).  This limit is designed to ensure a consistent solution structure from one time step to the next and is implemented as inner steps at a dynamically recalculated time step.   Using the PTL was shown to improve the solution for both the RKG2 STS scheme and for the implicit BE+PCG scheme.   Using the PTL reduces performance of the operators, in some cases by a large factor. However, it was found that the RKG2's original performance and scaling advantage over the BE+PCG scheme allow it to be competitive against the BE+PCG scheme even when using the PTL, all while exhibiting a much better solution than the original RKG2.  The results presented imply the following answers to the questions in Sec.~\ref{sec:results}:
\begin{enumerate}
    \item Can using the PTL with the RKG2 scheme obtain a solution similar (or better) to the original BE+PCG scheme? \emph{\color{OliveGreen} \\
    Yes, as shown in Figs.~\ref{fig:test1sol} and \ref{fig:test2sol}}.
    \item Does using the PTL with the BE+PCG scheme improve \emph{its} solution? \\ \emph{\color{OliveGreen} Yes, as shown in Figs.~\ref{fig:test1sol} and \ref{fig:test2sol}}.
    \item How does the PTL affect performance for both BE+PCG and RKG2 schemes? \\ \emph{\color{OliveGreen} In both cases, the performance decreases, however due to the original performance advantage of the RKG2 scheme, the PTL applied to RKG2 exhibits similar or better performance than the original BE+PCG.}
    \item Does using the PTL allow RKG2 to be competitive with the original BE+PCG scheme in both solution quality and performance? \\
    \emph{\color{OliveGreen} Yes, as shown in Figs.~\ref{fig:test1sol},\ref{fig:test2sol},\ref{fig:perf1} and \ref{fig:perf3}}
\end{enumerate}
We note that these answers are implied by the results obtained here, and are not guaranteed to be true in all cases.   We have tested the PTL in other models, such as our flux evolution code HipFT\footnote{Soon to be released at \url{https://www.github.com/predsci/hipft}
}, and our magnetic map smoothing tool called Diffuse \cite{DIFFUSEDC}.   In both these models, the PTL always avoided solution artifacts that were shown to occur when using very large time steps, with very little decline in performance.  Work is proceeding on the theoretical formulation and analysis of the PTL, including careful testing, especially in nonlinear cases \cite{PTLTHEORY}.  While the performance when applied to real-world simulations will be model and scheme dependent, the results shown here are encouraging that using the PTL can be used as a simple high performance way to avoid unwanted behavior of parabolic operators when using large time steps.  The PTL also may be a great way to ensure proper behavior of STS methods, making them a more robust and practical option for the integration of parabolic operators in multi physics models such as thermodynamic MHD.

\appendix

\section{}
\label{app:rkg2}
The RKG2 scheme with $\alpha=3/2$ \cite{o2019runge,skaras2021super} is given by 
\begin{equation}
\begin{array}{l}
u_0=u^n \\
y_0=F(u_0) \\
u_1 = u_0 + \tilde{\mu_1}\,\Delta t\,y_0  \\
\; \\
{\tt do\; k=2:s} \\
\;\;\;\; u_k = \mu_k\,u_{k-1} 
+ \nu_k\,u_{k-2} \\
\;\;\;\; \;\;\;\; + (1-\mu_k-\nu_k)\,u_0 \\
\;\;\;\; \;\;\;\; + \tilde{\mu_k}\,\Delta t\,F(u_{k-1}) 
+ \gamma_k\,\Delta t\,y_0 \\
{\tt enddo} \\
\; \\
u^{n+1} = u_s, 
\end{array}
\qquad
\vline
\qquad
\begin{array}{l}
{\tt do\; k=3:s} \\
\;\;\;\; b_k=\dfrac{4\,(k-1)\,(k+4)}{3\,k\,(k+1)\,(k+2)\,(k+3)} \\
\;\;\;\; \mu_k=\left(2 + \dfrac{1}{k}\right)\,\dfrac{b_k}{b_{k-1}} \\
\;\;\;\; \nu_k= -\left(\dfrac{1}{k} + 1\right)\,\dfrac{b_k}{b_{k-2}} \\
\;\;\;\; \tilde{\mu_k} = \mu_k\,w \\
\;\;\;\; \gamma_k=\left(\dfrac{k\,(k + 1)}{2}\,b_{k-1}-1\right)\,\tilde{\mu_k}, \\
{\tt enddo} \\
\end{array}
\notag
\end{equation}
where $u^n$ is the current solution, $u^{n+1}$ is the solution after integrating the full time step $\Delta t$, $F$ is the parabolic operator, $u_k$ is the solution at STS iteration $k$, and
\[
\begin{array}{l}
w=\dfrac{6}{(s+4)\,(s-1)}, \qquad b_0=1, \; b_1=\frac{1}{3}, \; b_2=\frac{1}{15}, \qquad \mu_1=1,
\\
\tilde{\mu_1}=w, \qquad \mu_2 = \dfrac{1}{2}, \qquad \nu_2 = -\dfrac{1}{10}, \qquad \tilde{\mu_2} = \mu_2\,w, \qquad \gamma_2 = 0. \\
\end{array}
\]
The number of required iterations $s$ is given by
\[
s=\left\lceil\frac{1}{2}\,\sqrt{25+24\,\frac{\Delta t}{\Delta t_{\mbox{\tiny Euler}}}}-\frac{3}{2}\right\rceil,
\]
where $\Delta t_{\mbox{\tiny Euler}}$ is the explicit Euler stability time step.  We note that it is critically important to ensure that the number of iterations be odd (by adding an iteration if needed), otherwise the amplification factor approaches one at the highest modes.

\section{}
\label{app:comp}
The details of the hardware and software specifications used in this paper are shown in Table~\ref{tab:sysinfo}.
\begin{table}[htbp]
\centering
\scalebox{0.9}{
$\begin{array}{|l|r|r|r|}
\multicolumn{1}{c}{\;} & 
\multicolumn{1}{c}{\bf SDSC\,Expanse} & 
\multicolumn{1}{c}{\bf NCSA\,Delta} &
\multicolumn{1}{c}{\bf Local\,GPU} \\
\multicolumn{1}{c}{\;} & 
\multicolumn{1}{c}{\bf CPU\,Node} & 
\multicolumn{1}{c}{\bf GPU(8x)\, Node} &
\multicolumn{1}{c}{\bf Workstation} \\
\hline
\mbox{CPU} & \mbox{(2x) AMD EPYC} & \mbox{(2x) AMD EPYC} & \mbox{Intel Core-i5} \\
\,& \mbox{Rome 7742} & \mbox{Milan 7763} & \mbox{Raptor Lake 13600KF} \\ \hline
\mbox{GPU} & \mbox{N/A} & \mbox{(8x) NVIDIA A100-40GB-SXM4} & \mbox{NVIDIA RTX 3090 Ti}\\ \hline
\mbox{CPU Cores} & 128 & 128 & 14\\  \hline
\mbox{DRAM} & 128\,\mbox{GB} & \mbox{320\,\mbox{GB}}\,\mbox{(GPUs)} & \mbox{24\,\mbox{GB}}\,\mbox{(GPU)}\\ \hline
 \mbox{Peak Memory} & \, &\, &\, \\
 \mbox{Bandwidth} 
& \mbox{410 GB/s} & \mbox{12,440 GB/s}\,\mbox{(GPUs)} & \mbox{1,008 GB/s}\,\mbox{(GPUs)} \\ \hline
\mbox{Max Flops} & \mbox{9 TFlops} & \mbox{78 TFlops}\,\mbox{(GPUs)} & \mbox{0.625 TFlops}\,\mbox{(GPUs)}\\ \hline
\mbox{Network} & \mbox{HDR InfiniBand} & \mbox{HPE SlingShot} & \mbox{N/A}\\ \hline
\mbox{MPI Library} & \mbox{OpenMPI 4.1.3} & \mbox{OpenMPI 3.1.5} & \mbox{OpenMPI 3.1.5}\\ \hline
\mbox{Compiler} & \mbox{GCC 10.2.0} & \mbox{NV HPC SDK 23.5} & \mbox{NV HPC SDK 23.5}\\ \hline
\mbox{OS} & \mbox{Rocky 8.7} & \mbox{Red Hat Enterprise 8.8} & \mbox{Mint 21.2}\\ \hline
\end{array}$}
\caption{System hardware and software configuration of SDSC Expanse, NCSA Delta, and our local workstation used for the results shown in this paper.\label{tab:sysinfo}}
\end{table}

\section*{Acknowledgments}
Work at Predictive Science Inc. was supported by the NASA LWS Strategic Capabilities Program (grant 80NSSC22K0893), the NSF PREEVENTS program (grant ICER1854790), and the NSF/NASA SWQU program (grants AGS 2028154 and 80NSSC20K1582).  
CDJ and LKSD acknowledge support from the NASA GSFC Heliophysics Internal Scientist Funding Model program (PI Jim Klimchuk).  CDJ also acknowledges support from the NASA LWS Focused Science Topic programs: NNH21ZDA001N-LWS ``The Origin of the Photospheric Magnetic Field: Mapping Currents in the Chromosphere and Corona'' (PI Pete Schuck) and NNH17ZDA001N-LWS ``Investigating Magnetic Flux Emergence with Modeling and Observations to Understand the Onset of Major Solar Eruptions'' (PI Mark Linton).
Computational resources at the San Diego Supercomputer Center and the National Center for Supercomputing Applications were provided by the NSF-supported ACCESS program (allocation MCA03S014).  We thank Puget Systems for providing the RTX 3090 Ti GPU for our local workstation. 

\section*{References}
\bibliographystyle{iopart-num}
\bibliography{main}

\end{document}